\documentclass[reprint,
showpacs,preprintnumbers,
amsmath,amssymb,
aps,
prb,
]{revtex4-1}

\usepackage{graphicx,appendix}
\usepackage{dcolumn}% Align table columns on decimal point
\usepackage{bm}% bold math
\usepackage{subfigure}
\usepackage[]{txfonts}
\usepackage{color}   %May be necessary if you want to color links
\usepackage{hyperref}
\hypersetup{
    colorlinks=true, %set true if you want colored links
    linktoc=all,     %set to all if you want both sections and subsections linked
    linkcolor=blue,  %choose some color if you want links to stand out
}

\begin{document}

\preprint{}

\title{Effects of Spin-Orbit coupling on Zero Energy Bound States Localized at Magnetic Impurities in Multi-Band Superconductors}

\author{Kangjun Seo$^1$}
\author{Jay D. Sau$^2$}
\author{Sumanta Tewari$^1$}

\affiliation{ $^1$Department of Physics and Astronomy, Clemson University, Clemson, SC 29634\\
$^2$Condensed Matter Theory Center and Joint Quantum Institute, Department of Physics,
University of Maryland, College Park, Maryland 20742-4111}

\date{\today}

\begin{abstract}
We investigate the effect of spin-orbit coupling on the in-gap bound states localized at magnetic impurities in multi-band superconductors with  unconventional (sign-changed) and conventional (sign-unchanged) $s$-wave pairing symmetry, which may be relevant to iron-based superconductors. Without spin-orbit coupling, for spin-singlet superconductors it is known that such bound states cross zero energy at a critical value of the impurity scattering strength and acquire a finite spin-polarization. Moreover, the degenerate, spin-polarized, zero energy bound states are unstable to applied Zeeman fields as well as deviation of the impurity scattering strength away from criticality. Using a T-matrix formalism as well as analytical arguments, we show that, in the presence of spin-orbit coupling, the zero-energy bound states localized at magnetic impurities in unconventional, sign-changed, $s$-wave superconductors acquire surprising robustness to applied Zeeman fields and variation in the impurity scattering strength, an effect which is absent in the conventional, sign-unchanged, $s$-wave superconductors. Given that the iron-based multi-band superconductors may possess a substantial spin-orbit coupling as seen in recent experiments, our results may provide one possible explanation to the recent observation of surprisingly robust zero bias scanning tunneling microscope peaks localized at magnetic impurities in iron-based superconductors provided the order parameter symmetry is sign changing $s_{+-}$-wave.
\end{abstract}


\maketitle

%\tableofcontents{}
%=======================================================
\section{Introduction}
We are motivated by a recent scanning tunneling microscope (STM) observation of a robust zero bias conductance peak (un-split by an applied magnetic field $\sim 8 T$) induced at magnetic impurities in iron-based superconductor Fe$_{1+x}$(Te,Se)~\cite{Yin2015}. Zero-- or low-energy sub-gap states bound to magnetic and/or non-magnetic impurities in superconductors are not unusual~\cite{Balatsky-2006,Yazdani-1997,Balatsky-1995,Tsai-2009,Bang-2009,Li-2009,Akbari-2010,Ng-2009,Matsumoto-2009,Zhang-2009,Kariyado-2010,Beaird-2012}. However, in spin-singlet superconductors [e.g., $s$-wave ($s_{++},s_{+-}$-wave), $d$-wave, etc], the zero bias peaks localized at impurities are expected to split into a double peak structure on application of a magnetic field. Qualitatively, the splitting of the peak by a magnetic field is due to the fundamental two-fold spin degeneracy of Bogoliubov quasi-particle states in a singlet superconductor. Since the spin of the zero-energy bound states couples to a magnetic field via Zeeman coupling, the zero bias STM peak, if any, gives rise to a double peak structure by application of a magnetic field. Theoretically, a defect-induced zero-energy state can escape splitting by a magnetic field when the state is non-degenerate. A non-degenerate zero energy bound state in a superconductor, on the other hand, is very unusual, and is commonly referred as a Majorana bound state (MBS) that can be realized in a topological superconductor~\cite{Read_Green_2000,Kitaev_2001}. This has led to the tantalizing conjecture of realizing a topological superconductor and MBS in iron-based superconductors induced by superconductivity, spin-orbit coupling, and local magnetic order induced at isolated magnetic impurities~\cite{Yin2015}.

 Impurity induced in-gap states at non-magnetic and magnetic impurities in iron-based superconductors have been investigated before within a T-matrix approach and the Bogoliubov de-Gennes formalism \cite{Tsai-2009,Bang-2009,Li-2009,Akbari-2010,Ng-2009,Matsumoto-2009,Zhang-2009,Kariyado-2010,Beaird-2012}. In both approaches it has been found that, while for non-magnetic impurities in-gap bound states exist only for unconventional, sign-changed, $s$-wave superconductors ($s_{+-}$), for magnetic impurities such states exist for both sign-unchanged ($s_{++}$) and sign-changed ($s_{+-}$) superconducting ordering symmetries. For magnetic impurities, with increasing strength of the impurity potential, a pair of spin-polarized in-gap bound states cross zero energy at a quantum phase transition at a critical value of the scattering potential. Moreover, the pair of zero energy bound states at the critical scattering strength, owing to finite spin-polarizations, are unstable to applied Zeeman fields, which split them into a pair of positive and negative energy in-gap states, producing a double peak structure in tunneling experiments. Thus, within this conventional picture of impurity scattering in iron-based superconductors \cite{Tsai-2009,Bang-2009,Li-2009,Akbari-2010,Ng-2009,Matsumoto-2009,Zhang-2009,Kariyado-2010,Beaird-2012}, zero bias STM peaks      at magnetic impurities that remain unsplit by magnetic fields $\sim 8 T$ cannot be explained, irrespective of whether the superconducting ordering symmetry is assumed to be $s_{+-}$ or $s_{++}$-wave. More recently, Ref.~[\onlinecite{Tai2015}] has attempted to explain the robustness of zero bias STM peaks at magnetic impurities in iron-based superconductor Fe$_{1+x}$(Te,Se)~\cite{Yin2015} in terms of a $Z_2$ topological mirror order and $s_{+-}$ superconducting order symmetry. In other recent work, Ref.~[\onlinecite{Huang2016}] has attempted to explain the same experiments ~\cite{Yin2015} within a so-called `tunneling impurity' formulation, in which the magnetic impurity is assumed to be coupled to the underlying Fe lattice only by hoping terms, but no exchange interaction, in spite of the fact that the impurity possesses a non-zero local magnetic moment. In this paper we show that the zero-energy bound states localized at magnetic impurities in sign changing $s_{+-}$-wave superconductors (but not in sign unchanged $s_{++}$-wave superconductors) can be surprisingly robust to perturbations such as Zeeman fields and variations in the impurity scattering strength in the presence of spin-orbit coupling. Given that a substantial spin-orbit coupling ($\sim 5-10$ meV) may be present in all the classes of iron based superconductors as seen in recent experiments \cite{Borisenko2016}, our work provides an alternative explanation of robust STM peaks in iron-based superconductor Fe$_{1+x}$(Te,Se)~\cite{Yin2015} which remain unsplit even by a magnetic field as high as $\sim 8 T$  without having to invoke exotic physics such as topological superconductivity \cite{Tai2015,Yin2015} and/or absence of exchange coupling of magnetic impurity with the underlying Fe lattice \cite{Huang2016}. In addition, our work fills the gap of analyzing the effects of spin-orbit coupling on magnetic impurity induced Yu-Shiba-Rusinov (YSR) states \cite{Yu-1965,Shiba-1968,Rusinov-1969} in unconventional and conventional $s$-wave superconductors.
  
  In our calculations, the key to the robustness of magnetic impurity induced zero energy bound states in multi-band unconventional $s$-wave superconductors is a non-zero spin-orbit coupling (SOC), which has so far been neglected in the analysis of YSR states in multi-band superconductors. In recent high resolution ARPES experiments, substantial SOC ($\sim 5-10$ meV) has been detected in all the families of iron-based superconductors via the observation of SOC-induced Fermi surface splitting~\cite{Borisenko2016}.
 %The low-energy sub-gap states at the impurity sites are either a consequence of SOC (for large enough SOC, the system is a nodal superconductor, see Fig. 2 (d,e,f)), or may naturally arise from overlap of the impurity-induced sub-gap states bound to nearby magnetic impurities. We note that the local density of states (LDOS) measured at the impurity sites in samples with a finite density of magnetic impurities indeed indicates a finite density of low-energy sub-gap quasiparticle states, in addition to a robust zero bias peak un-split by a magnetic field \cite{Yin2015}.
 With a minimal modeling of the band Hamiltonian in the presence of SOC we are able to show that a robust zero bias state (ZBS) is induced at magnetic impurities in iron-based superconductors provided the symmetry of the superconducting order parameter is sign-changing $s_{+-}$-wave.
  The robustness of the ZBS due to SOC can be intuitively understood as resulting from the suppression of the superconducting gap by SOC in the $s_{+-}$ case and the properties of impurities in nearly isotropic gap superconductors \cite{Yu-1965,Shiba-1968,Rusinov-1969}. As shown in Fig.~\ref{fig:fig1}, SOC produces a finite Fermi surface splitting that pushes one Fermi surface towards the X-point and the other towards the $\Gamma$-point of the Brillouin zone. On the other hand, for $s_{+-}$ pairing, the pairing potential changes sign between these two points in the Brillouin zone. Therefore, with one Fermi surface moving towards the $X$-point, the pairing amplitude on that Fermi surface is suppressed. As a result, one of the Fermi surfaces in $s_{+-}$-wave superconductors in the presence of SOC would have a smaller gap. This suppressed gap is still quite isotropic and as shown in the Appendix, for such isotropic Fermi surfaces, one can prove that a magnetic impurity will support localized sub-gap bound states just as in the case of the Yu-Shiba-Rusinov states \cite{Yu-1965,Shiba-1968,Rusinov-1969}. However, if the gap is suppressed by SOC, then the impurity-induced sub-gap state is pinned to live inside the smaller gap, even in the presence of a magnetic field, explaining the robustness of the zero bias peak to substantial magnetic fields.
 On the other hand, the magnitude of conventional $s_{++}$ pairing gap does not renormalize in the presence of SOC. Thus, the $s_{++}$ superconductor does not provide the robust zero-energy bound states at magnetic impurities, with zero bias states being strongly affected by the perturbation of a magnetic field even with spin-orbit coupling. Thus, our results, in addition to providing a possible theoretical explanation of robust zero bias conductance peaks in iron-based superconductors~\cite{Yin2015} without having to invoke exotic physics such as topological superconductivity \cite{Tai2015,Yin2015} or absence of exchange coupling between magnetic impurity and the underlying Fe lattice \cite{Huang2016}, also helps in identifying the relevant symmetry of the superconducting order parameter of iron-based superconductors as sign-changing $s_{+-}$-wave.

The paper is organized as follows: In Sec.~\ref{sec2}, we introduce the model Hamiltonian and the formalism. The robustness of the zero-energy bound states induced by a single magnetic impurity with and without spin orbit coupling is investigated in Sec.~\ref{sec31}. Then, we present the effects of the multiple magnetic impurities in Sec.~\ref{sec32}. Finally, a conclusion is given in Sec.~\ref{sec4}. Some technical details pertaining to the analytical calculations of the robustness are relegated to the Appendix.

\section{Model and formalism}\label{sec2}
We start with a mean-field Hamiltonian for the iron-based superconductor using the two-orbital model ($d_{xz}$ and $d_{yz}$) on the two-dimensional Fe square lattice~\cite{Raghu-2008},
\begin{equation}
	H = H_0 + H_\text{mag} + H_\text{imp}.
\end{equation}
Here, $H_0$ is the tight-binding Hamiltonian in the superconducting state, including intra- and inter-orbital hopping integrals
\begin{equation}
	H_0=\sum_{\mathbf{i}\mathbf{j}\alpha\beta\sigma} t_{\mathbf{i}\mathbf{j}}^{\alpha\beta} c_{\mathbf{i}\alpha\sigma}^\dag c_{\mathbf{j}\beta\sigma} - \mu \sum_{\mathbf{i}\alpha\sigma} c_{\mathbf{i}\alpha\sigma}^\dag c_{\mathbf{i}\alpha\sigma}
	 + H_\text{pair},
\end{equation}
where $c_{\mathbf{i}\alpha\sigma}^\dag$ creates an electron with spin $\sigma$ in the orbitals $\alpha=1$ ($d_{xz}$) and $2$ ($d_{yz}$) at site $\mathbf{i}$. Following earlier work ~\cite{Tsai-2009}, we take the values of the nearest-neighbor hopping matrix elements as
$t_{\mathbf{i}\pm\hat{x}}^{11}=t_{\mathbf{i}\pm\hat{y}}^{22}=t_1$,
$t_{\mathbf{i}\pm\hat{x}}^{22}=t_{\mathbf{i}\pm\hat{y}}^{11}=t_2=-1.3t_1$, and the next-nearest-neighbor hopping as
$t_{\mathbf{i}\pm(\hat{x}+\hat{y})}^{\alpha\alpha}=t_{\mathbf{i}\pm(\hat{x}-\hat{y})}^{\alpha\alpha}=t_3$, and $t_{\mathbf{i}\pm(\hat{x}+\hat{y})}^{\alpha\beta}=-t_{\mathbf{i}\pm(\hat{x}-\hat{y})}^{\alpha\beta}=t_4$  with $t_3 = t_4=0.85t_1$. We have taken $t_1=10$ meV as the energy units and lattice constant $a= 1$. The chemical potential $\mu=1.65 t_1$ is adjusted to give a fixed filling factor $n_e \simeq 2.1$ per site~\cite{Tsai-2009}. We have checked that the main result of this paper -- robust zero energy states at magnetic impurities in $s_{+-}$ superconductors in the presence of spin-orbit coupling -- is robust to variations in these parameters as long as the superconducting order parameter symmetry is sign changing $s_{+-}$-wave. 
%============================================
\begin{figure}[t]
\includegraphics[width=1.0\linewidth]{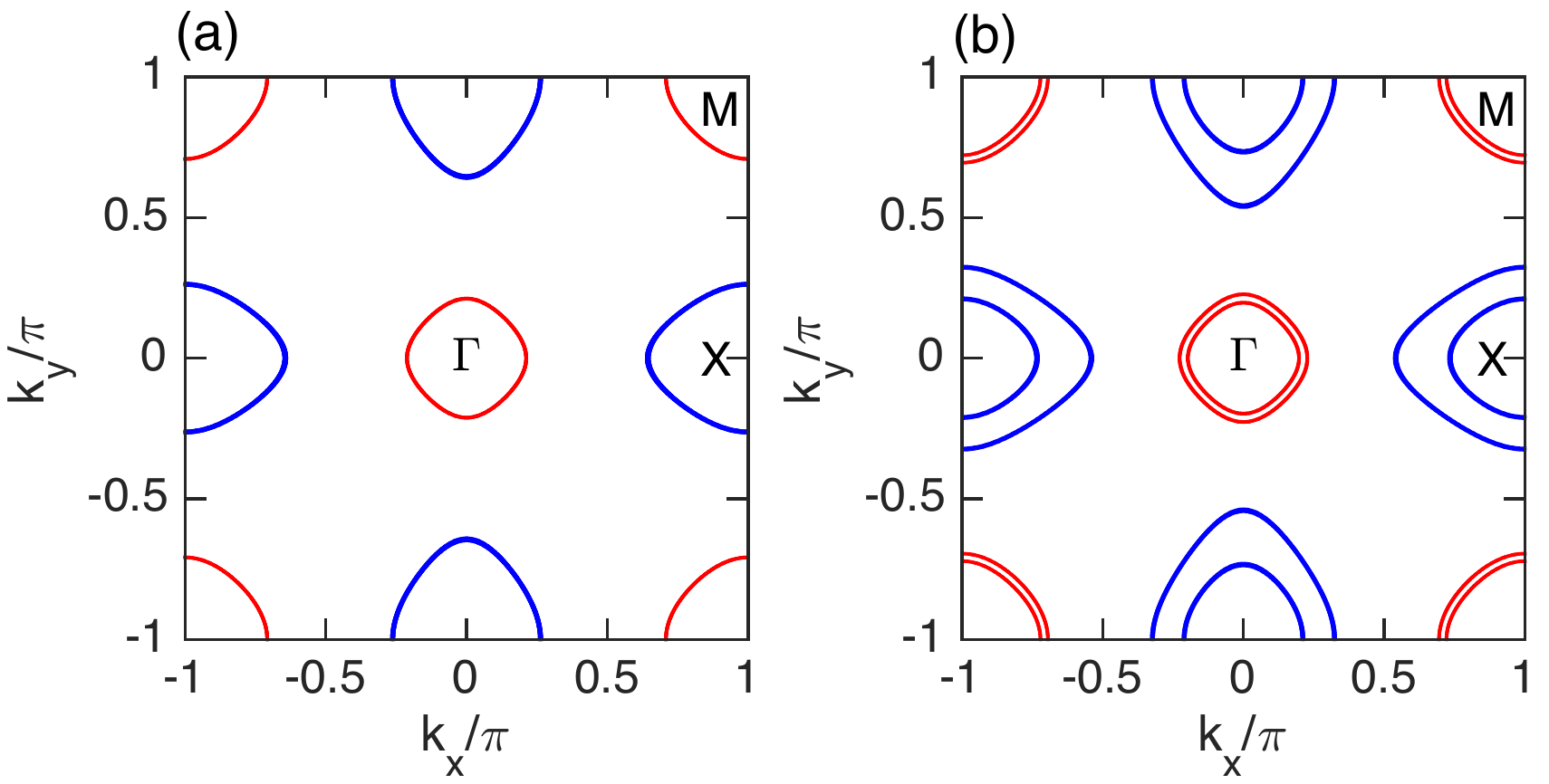}
\caption{
(a) Fermi surfaces of the two-band model at $\mu=1.65 t_1$ in the unfolded BZ, and (b) the fermi surfaces of the helicity bands in the presence of the Rashba-type SOC with $\nu = 0.25 t_1 = 2.5$ meV. The red pockets at $\Gamma$ and $M$ points represent the holelike pockets, and the blue pockets at $X$ point represent the  electronlike pockets. SOC produces the additional Fermi surfaces with the same chemical potential.
}
\label{fig:fig1}
\end{figure}
%============================================
The Fermi surfaces consist of the hole pockets at $\Gamma$ and $M$ points and the electron pocket at $X$ point in the unfolded Brillouin Zone [Fig.~\ref{fig:fig1}(a)] without SOC. 

The pairing Hamiltonian $H_\text{pair}$ is given by
\begin{equation}
	H_\text{pair}=\sum_{\mathbf{i}\mathbf{j}\alpha} \Delta_\alpha (\mathbf{i},\mathbf{j}) c_{\mathbf{i}\alpha\uparrow}^\dag c_{\mathbf{j}\alpha\downarrow}^\dag +h.c.,
\end{equation}
where $\Delta_\alpha(\mathbf{i},\mathbf{j})$ is a mean-field superconducting order parameter. We focus on the unconventional sign-changed $s_{+-}$ pairing symmetry so that $\Delta_\alpha (\mathbf{i},\mathbf{j})=\Delta_0\delta_{\mathbf{i},\mathbf{j}\pm(\hat{x}\pm\hat{y})}$~\cite{kseo-2008,Tsai-2009}. For a conventional $s_{++}$ superconductor, $\Delta_\alpha (\mathbf{i},\mathbf{j}) = \Delta_0 \delta_{\mathbf{i},\mathbf{j}}$.

We can rewrite the Hamiltonian in the momentum space as $H_0=\frac{1}{N}\sum_\mathbf{k} \Psi_\mathbf{k}^\dag \hat{h}_\mathbf{k}^0 \Psi_\mathbf{k}$, where
\begin{equation}
\hat{h}_\mathbf{k}^0
	=
\xi^i_\mathbf{k} \hat\alpha^i \otimes (\hat\tau^3\otimes \hat\sigma^0)
+ \Delta_\mathbf{k} \hat\alpha^0\otimes(\hat\tau^1\otimes\hat\sigma^0),
\end{equation}
and $\Psi_\mathbf{k}$ is an 8-dimensional Nambu spinor
$\Psi_{\mathbf{k}}^\dag
=
[c_{\mathbf{k} 1\uparrow}^\dag,c_{\mathbf{k}1\downarrow}^\dag,c_{-\mathbf{k}1\downarrow},-c_{-\mathbf{k}1\uparrow},
c_{\mathbf{k}2\uparrow}^\dag,c_{\mathbf{k}2\downarrow}^\dag,c_{-\mathbf{k}2\downarrow},-c_{-\mathbf{k}2\uparrow}]$ with $c_{\mathbf{k}\alpha\sigma}$ being a Fourier transform of $c_{\mathbf{r}\alpha\sigma}$. The Pauli matrices $\hat\alpha^i$, $\hat\sigma^i$ and $\hat\tau^i$ act on the orbital, the particle-hole and the spin spaces, respectively. Then, the order parameter $\Delta_\mathbf{k} = \Delta_0 \cos k_x \cos k_y$ for $s_{+-}$ pairing, and the dispersions
$\xi_\mathbf{k}^0=(t_1+t_2) (\cos k_x + \cos k_y)+4t_3 \cos k_x \cos k_y-\mu$,
$\xi_\mathbf{k}^1=4t_3 \sin k_x \sin k_y$,
$\xi_\mathbf{k}^2=0$, and
$\xi_\mathbf{k}^3=(t_1-t_2) (\cos k_x - \cos k_y)$.

The second term in the Hamiltonian $H_\text{mag}$ describes the effects of the magnetic field and SOC. In our work, we consider the out-of-plane magnetic field $h_\text{ext}$, and the Rashba-type SOC~\cite{rashba-1984} with an angular momentum $\mathbf{L}(\mathbf{k}) = (-\sin k_y , \sin k_x ,0)$~\cite{rashba-2001,Frigeri-2004} on the two-dimensional $x-y$ plane. Then, we have $H_\text{mag}=\frac{1}{N}\sum_\mathbf{k} \Psi_\mathbf{k}^\dag \hat{h}_\mathbf{k}^\text{mag} \Psi_\mathbf{k}$, where
\begin{equation}
	\hat{h}_\mathbf{k}^\text{mag}=
	-h_\text{ext} \hat\alpha^0\otimes(\hat\tau^0\otimes\hat\sigma^3)
	+
	\nu \hat\alpha^0\otimes(\hat\tau^3 \otimes \mathbf{L}(\mathbf{k}) \cdot \bm{\hat\sigma})
\end{equation}
with $\nu$ being the strength of SOC.

In this work, we treat the magnetic impurity as a localized spin in the classical limit ($S\gg 1$)~\cite{Yu-1965,Shiba-1968,Rusinov-1969,sau-2013}, and the quantum (Kondo) effect of impurity is not under our consideration. In this limit, which has been studied earlier quite extensively for $s_{+-}$ and $s_{++}$ superconductors in the absence of spin-orbit coupling \cite{Tsai-2009,Bang-2009,Li-2009,Akbari-2010,Ng-2009,Matsumoto-2009,Zhang-2009,Kariyado-2010,Beaird-2012}, the magnetic impurity is equivalent to the local magnetic moment $\mathbf{S}$. Then the impurity Hamiltonian describes the interaction between the conduction electrons and the impurity spin located at $\mathbf{r} =0$
\begin{equation}
	H_\text{imp}=\sum_{\alpha} \mathbf{S} \cdot (J_1 \mathbf{s}_{\alpha\alpha} (0)+J_2 \mathbf{s}_{\alpha\bar{\alpha}} (0)),
\end{equation}
where $J_1$ and $J_2$ are the intra- and inter-orbital exchange couplings, and the operators $\mathbf{s}_{\alpha\beta}(\mathbf{r})=\frac{1}{2}\sum_{\sigma\sigma'} c_{\mathbf{r}\alpha\sigma}^\dag \hat \tau_{\sigma\sigma'} c_{\mathbf{r}\beta\sigma'}$. Spin-rotational symmetry of the system enables us to choose the $z$ axis of the spin degrees of freedom to point in the direction of $\mathbf{S}$.  Using the Nambu spinor $\Psi_\mathbf{k}$, the impurity Hamiltonian can be rewritten as $H_\text{imp}=\sum_{\mathbf{k},\mathbf{k}^\prime} \Psi_\mathbf{k}^\dag \hat V \Psi_{\mathbf{k}^\prime}$ with $\hat V = J_1 S \hat\alpha^0\otimes (\hat\tau^0 \otimes \hat\sigma^3) + J_2 S \hat\alpha^1 \otimes ( \hat\tau^0\otimes \hat\sigma^3)$. We shall consider the effects of the intra-orbital impurity scattering, thus the strength of the impurity is given by $w=S_z J_1$.

We perform a numerical study employing a mean-field $T$-matrix approximation~\cite{Balatsky-2006,Pientka-2013}. In this case, we assume that the spatial variation of the superconducting order parameter can be neglected. Since the impurity interaction is limited to one site, scattering of quasiparticles from the impurity moment is described by a $T$-matrix, $\hat T(\omega)$, whose Fourier transform is independent of wave vectors. Then the single-particle Green's function for an impurity located at $\mathbf{r}=0$ is given by
\begin{equation}
\label{green_func}
	\hat G(\mathbf{r},\mathbf{r}^\prime;\omega)
	=
	\hat G^{(0)}(\mathbf{r}-\mathbf{r}',\omega)
	+
	\hat G^{(0)}(\mathbf{r},\omega) \hat T(\omega) \hat G^{(0)}(-\mathbf{r}^\prime,\omega),
\end{equation}
where $\hat G^{(0)}(\mathbf{r},\omega)=\frac{1}{N}\sum_\mathbf{k} \hat G^{(0)}(\mathbf{k},\omega) e^{i\mathbf{k}\cdot \mathbf{r}}$. The single-particle Green's function for a clean system
$\hat G^{(0)}(\mathbf{k},\omega)=[(\omega + i0^+) I -( \hat{h}_\mathbf{k}^0 + \hat{h}_\mathbf{k}^\text{mag}) ]^{-1}$, where $I$ is a 8-dimensional identity matrix. The Fermi surfaces in the presence of SOC at zero magnetic field is presented in Fig.~\ref{fig:fig1}(b). The helicity bands remove the degeneracies of the electron spin in the electronlike and holelike pockets. With $\hat G^{(0)}(\mathbf{r},\omega)$ in hands, the $T$-matrix can be obtained from the Lippmann-Schwinger equation
\begin{equation}
	\hat T(\omega) = \left[ \hat I - \hat V \hat G^{(0)}(\mathbf{0},\omega) \right]^{-1} \hat V.
\label{eq:lippmann}
\end{equation}
Note that as far as the spatial variation of order parameter can be neglected, these equations allow a complete solution of the problem.

The nature of the magnetic impurity induced bound states can be found by computing the spin-resolved local density of states (LDOS)
\begin{equation}
	N_{\alpha\sigma}(\mathbf{r},\omega)=-\frac{1}{\pi}\textrm{Im }G_{\alpha\sigma,\alpha\sigma}(\mathbf{r},\mathbf{r};\omega)
\end{equation}
of which the poles give the energy spectra of single-particle excitations, and consist of those of the $\hat G_0$ and the $T$ matrix. The poles of $T$-matrix signifies the emergence of the impurity induced states. It is known that a strong scattering yields localized states deep in the gap, while a weak scattering results in bound states close to the gap edge~\cite{Balatsky-2006}.

For the case of a quantum spin, one needs to address the Kondo effect~\cite{Polkovnikov-2001,Zhang-2001,Zhu-2001}. However, following earlier work on $s$-wave superconductors including $s_{+-}$ and $s_{++}$ \cite{Tsai-2009,Bang-2009,Li-2009,Akbari-2010,Ng-2009,Matsumoto-2009,Zhang-2009,Kariyado-2010,Beaird-2012}, in this work, we treat the magnetic impurity as a classical spin using a mean-field $T$-matrix approximation approach. In this case the main effect of the exchange coupling between the local moment $S$ and electron spin is the renormalization of the effective scattering potential for electrons of two different spin orientations, and so there are four impurity induced in-gap states, one for each electron-spin orientation in each $d_{xz}$ ($\alpha=1$) and $d_{yz}$ ($\alpha=2$) orbitals. It is noteworthy that the degeneracy between two orbitals, $d_{xz}$ and $d_{yz}$, cannot be removed by the magnetic field and the SOC as well, thus we omit the notation of $\alpha$ in the spin-resolved LDOS throughout this paper.

%============================================
\begin{figure}[h]
\includegraphics[width=1.0\linewidth]{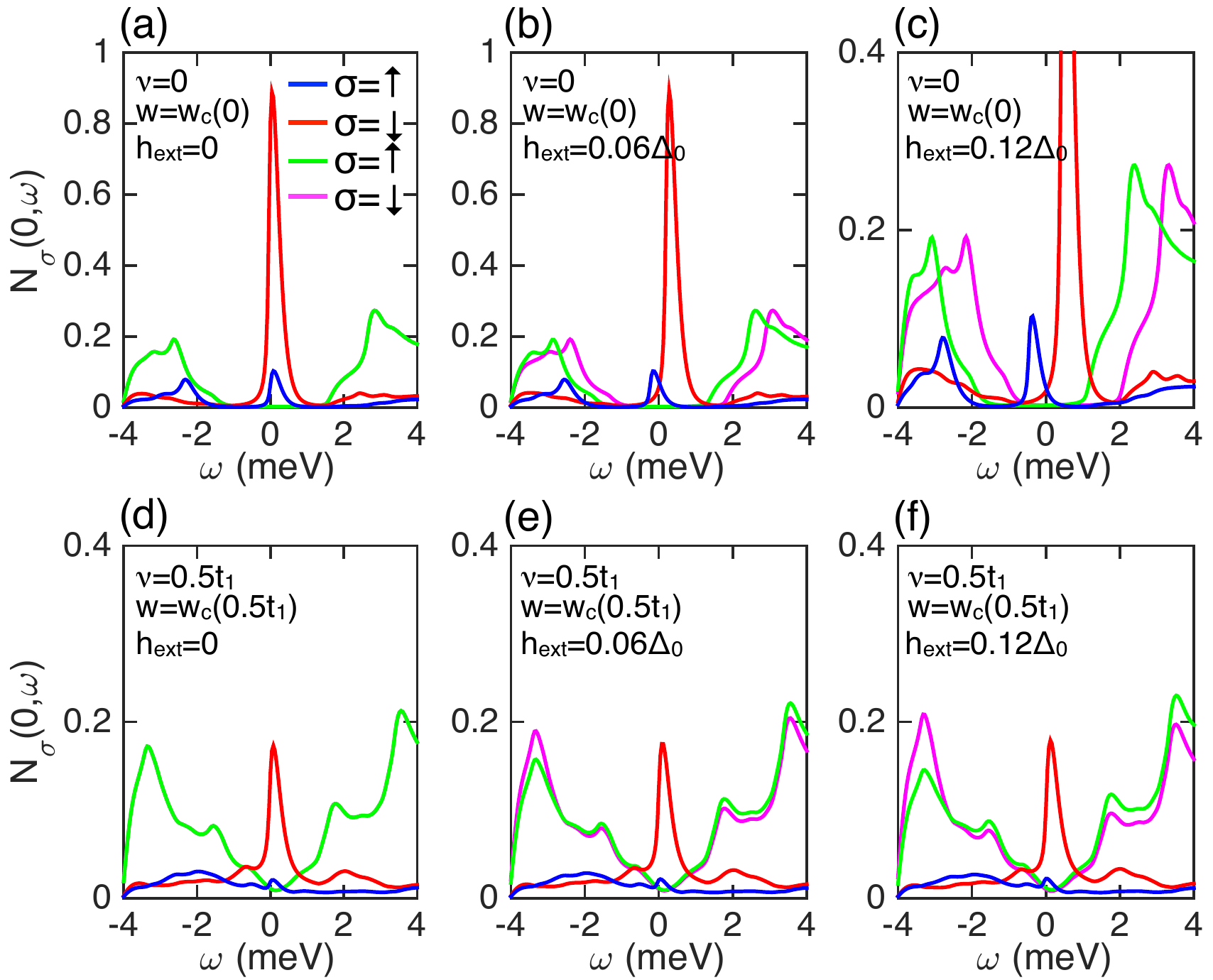}
\caption{
Magnetic field and SOC effects on LDOS, $N_\sigma$ for $s_{+-}$ pairing vs bias energy $\omega$ for clean system (green and magenta for $\sigma=\uparrow, \downarrow$, respectively) and system with an impurity at $\mathbf{r} = 0$ (blue and red for $\sigma=\uparrow, \downarrow$) at $T=1.5$ K. (a) and (d) show the appearance of both ZBPs for $\sigma=\uparrow$ and $\downarrow$ at the critical values $w=w_c(\nu=0)$ and $w=w_c(\nu=0.5t_1)$, respectively; (b) and (c) illustrate the Zeeman splitting of $N_{\uparrow}$ and $N_{\downarrow}$ at weak magnetic fields $h_\text{ext}=0.06\Delta_0$ and $0.12\Delta_0$ without SOC. Whereas (e) illustrates the robustness of the ZBPs, $N_{\uparrow,\downarrow}$, to the applied magnetic field in the presence of SOC, $\nu=0.5 t_1 = 5$ meV, (f) the Zeeman split begins to appear with increasing magnetic field for a single impurity in the system. Comparing the LDOS for clean systems (green and magenta), it is manifest that the SOC slightly changes the low energy states by forming a V-shaped LDOS, and make the system robust against magnetic field.
}
\label{fig:fig2}
\end{figure}
%============================================
\section{Single magnetic impurity}\label{sec31}
We begin with the effects of the applied magnetic field and SOC on the in-gap bound states induced by a single magnetic impurity for the sign changed $s_{+-}$-wave iron-based superconductor. Fig.~\ref{fig:fig2} presents the spin-resolved LDOS $N_{\sigma}(\mathbf{r}=0,\omega)$ at the critical values of impurity strengths $w=w_c(\nu=0)$ and $w=w_c(\nu=0.5t_1)$ for a finite temperature $T=1.5$ K below the superconducting critical temperature $T_c$. For a weak impurity scattering $w < w_c$, the ground state has time-reversed pairs of single-particle in-gap states $\pm \Omega$. As $w$ increases, the energies $\pm\Omega$ approach the chemical potential and, eventually, at the critical scattering strength $w=w_c$, it becomes a zero energy state, or zero-energy bound state (ZBS)~\cite{Balatsky-1995,Yazdani-1997,Balatsky-2006,Tsai-2009}. Fig.~\ref{fig:fig2}(a) and (d) show the ZBSs for both spins ($\sigma=\uparrow$ and $\sigma=\downarrow$) with zero magnetic field applied, $h_\text{ext}=0$. In the absence of SOC ($\nu = 0$), the corresponding zero bias peaks (ZBPs) in the LDOS at $h_\text{ext}=0$ begin to split with increasing magnetic field $h_\text{ext}$. Fig.~\ref{fig:fig2}(b) and (c) illustrate the Zeeman splitting by the applied magnetic fields $h_\text{ext}=0.06 \Delta_0$ ($\sim 4$ Tesla) and $h_\text{ext}=0.12 \Delta_0$ ($\sim 8$ Tesla), respectively. In contrast, the presence of SOC ($\nu=0.5t_1 \sim 5$ meV) dramatically reduces the Zeeman splitting and makes the ZBPs robust to the magnetic field [Fig.~\ref{fig:fig2}(e) and (f)]. Note that SOC $\sim 5-10$ meV may not be unrealistic in iron based superconductors as seen in recent high resolution ARPES experiments \cite{Borisenko2016}.  It is noteworthy that the ZBSs localized at the impurity site in the presence of SOC remain pinned to zero energy even in a magnetic field $0.12 \Delta_0 \sim 8$ Tesla. It is in good agreement with the experimental observations~\cite{Yin2015}. We believe that the robust ZBS is a strong signature of the presence of SOC in the system. Note, however, the low energy quasiparticle states inside the gap in the clean system (green and magenta) even at $h_\text{ext}=0$ [Fig.~\ref{fig:fig2}(d),(e),(f)]. They exist because a finite SOC reduces the magnitude of the superconducting gap, as discussed in the introduction, and with a finite SOC $\nu=0.5t_1=5$ meV the system is a nodal superconductor. As discussed in the introduction (also see below and the Appendix for more details), the reduction of the magnitude of the superconducting gap in $s_{+-}$ (but not in $s_{++}$) superconductors with SOC is the key effect responsible for the increased robustness of magnetic impurity induced ZBSs to applied Zeeman fields. Later we will show that in a system with a finite concentration of magnetic impurities as in the experiments \cite{Yin2015}, the low energy quasiparticle density which effectively reduces the gap can naturally arise from YSR states bound to the nearby impurities. In this case, the ZBSs localized at impurity sites are robust to applied magnetic fields (as well as to variations in the impurity potential) even for smaller values of SOC ($\nu=0.25t_1=2.5$ meV), corresponding to which the clean system has a full gap. This demonstrates that our results are robust and do not depend on the specific values of the parameters, as long as there is SOC and a finite concentration of low energy quasiparticles at the impurity sites as in the experiments.

The robustness of the ZBS due to SOC is attributed to a combined effets of suppression of the superconducting gap by SOC in the $s_{+-}$ case and the properties of magnetic impurities in nearly isotropically gapped superconductors. As seen in Fig.~\ref{fig:fig1}, SOC produces a relatively large spin splitting that pushes one Fermi surface towards the X-point and the other towards the $\Gamma$-point in the Brillouin zone. On the other hand, in $s_{+-}$ pairing, the pairing potential changes sign between these two points. Therefore, as one Fermi surface moves towards the $X$-point, the pairing amplitude is suppressed, so that one of the Fermi surfaces would have a smaller gap. This suppressed gap is still quite isotropic and as shown in the appendix, for such isotropic Fermi surfaces, one can prove that an infinitesimal magnetic impurity will support sub-gap states just as in the case of the YSR states \cite{Yu-1965,Shiba-1968,Rusinov-1969}. However, if the gap is suppressed by SOC, then the impurity-induced sub-gap state is pinned to live inside the smaller gap, even in the presence of a magnetic field, explaining the robustness of the ZBP to substantial magnetic fields. Since the coherence peaks from the larger gapped Fermi surfaces are expected to be larger, the smaller gap could appear as a pinned peak inside the larger gap.

%============================================
\begin{figure}[t]
\includegraphics[width=1.0\linewidth]{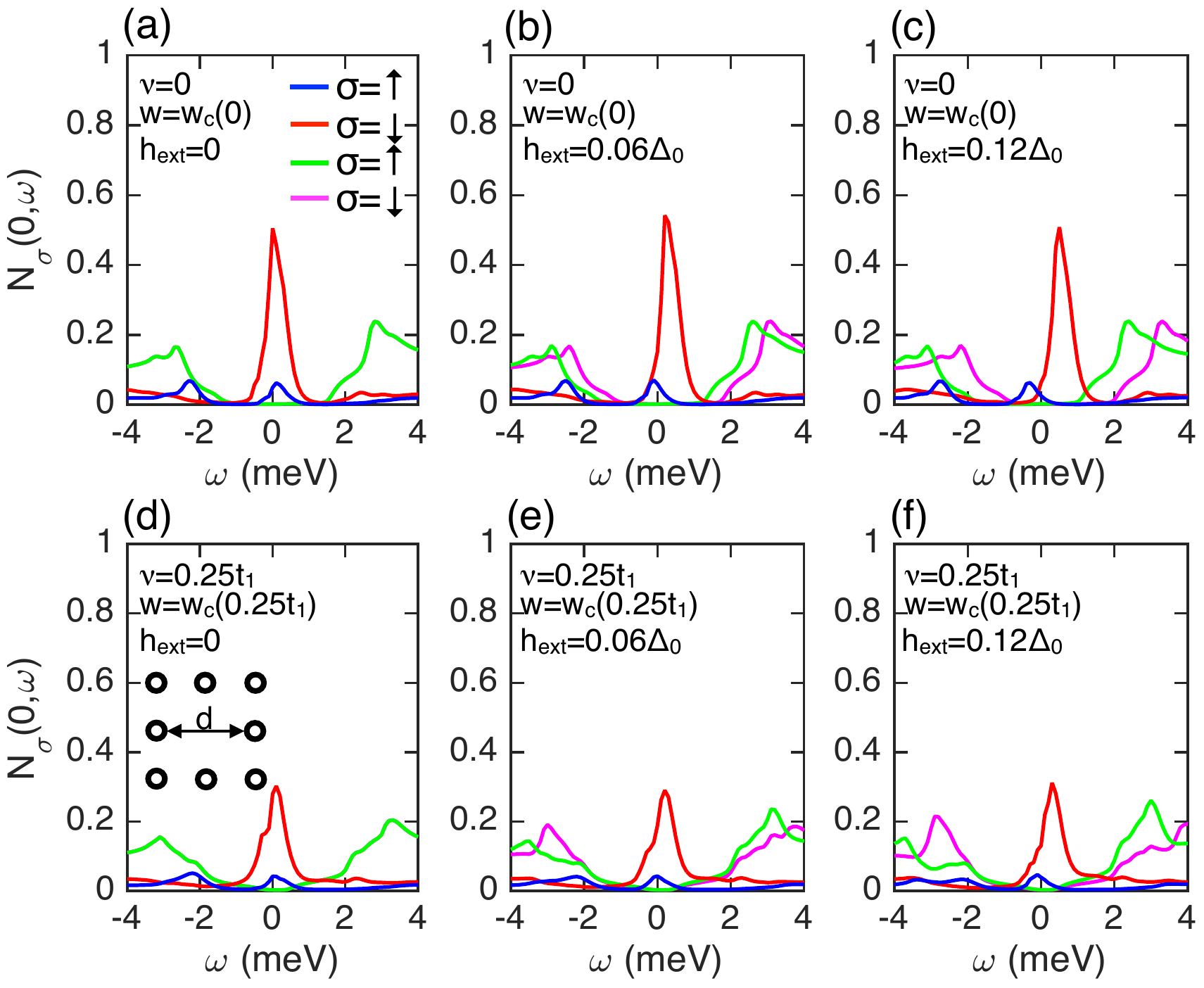}
\caption{
Magnetic field and SOC effects on LDOS, $N_\sigma$ for $s_{+-}$ pairing vs bias energy $\omega$ for clean system (green and magenta for $\sigma=\uparrow, \downarrow$, respectively) and system with multiple impurities (blue and red for $\sigma=\uparrow, \downarrow$) at $T=1.5$ K. 8 impurities are located in a square and $d =12$ atomic sites. The LDOS are calculated at a impurity site. (a)--(c) show the ZBPs for $\sigma=\uparrow$ and $\downarrow$ at the critical value value of the impurity potential without SOC, $w=w_c(\nu=0)$. The ZBP is split by applying magnetic field $h_\text{ext}=0.12\Delta_0$; (d)--(f) illustrate the robustness of the ZBPs, $N_{\uparrow,\downarrow}$ in the presence of SOC, $\nu=0.25 t_1 = 2.5$ meV.
}
\label{fig:fig3}
\end{figure}
%============================================
\section{Multiple magnetic impurities}\label{sec32}
Now we consider the robust ZBSs induced by the multiple magnetic impurities for $s_{+-}$-wave superconductor in the presence of SOC. This addresses the so-called class-D anti-localization mechanism. The case of a large number of bound states to magnetic impurities in a superconductor with SOC is described by random-matrix theory in symmetry class D~\cite{Altland}. This symmetry class shows no level repulsion at zero energy. Because of this, the disorder averaged density of states (or the density of states of a large number of weakly localized bound states) is expected to show a peak at zero energy quite generically independent of the magnetic field~\cite{Bagrets}. Specifically, in the present context, nearby multiple Shiba impurities provide a second mechanism (aside from SOC) of reducing the magnitude of local gap near a given magnetic impurity. By the argument given in the appendix, the zero bias peak at the given impurity will then stay pinned at zero energy even if the value of SOC is smaller, provided there is overlap of wave functions from states localized at nearby impurities. This demonstrates that our results are robust to specific values of SOC. Below we explore this mechanism numerically by introducing multiple Shiba impurities.

%============================================
\begin{figure}[b]
\includegraphics[width=1.0\linewidth]{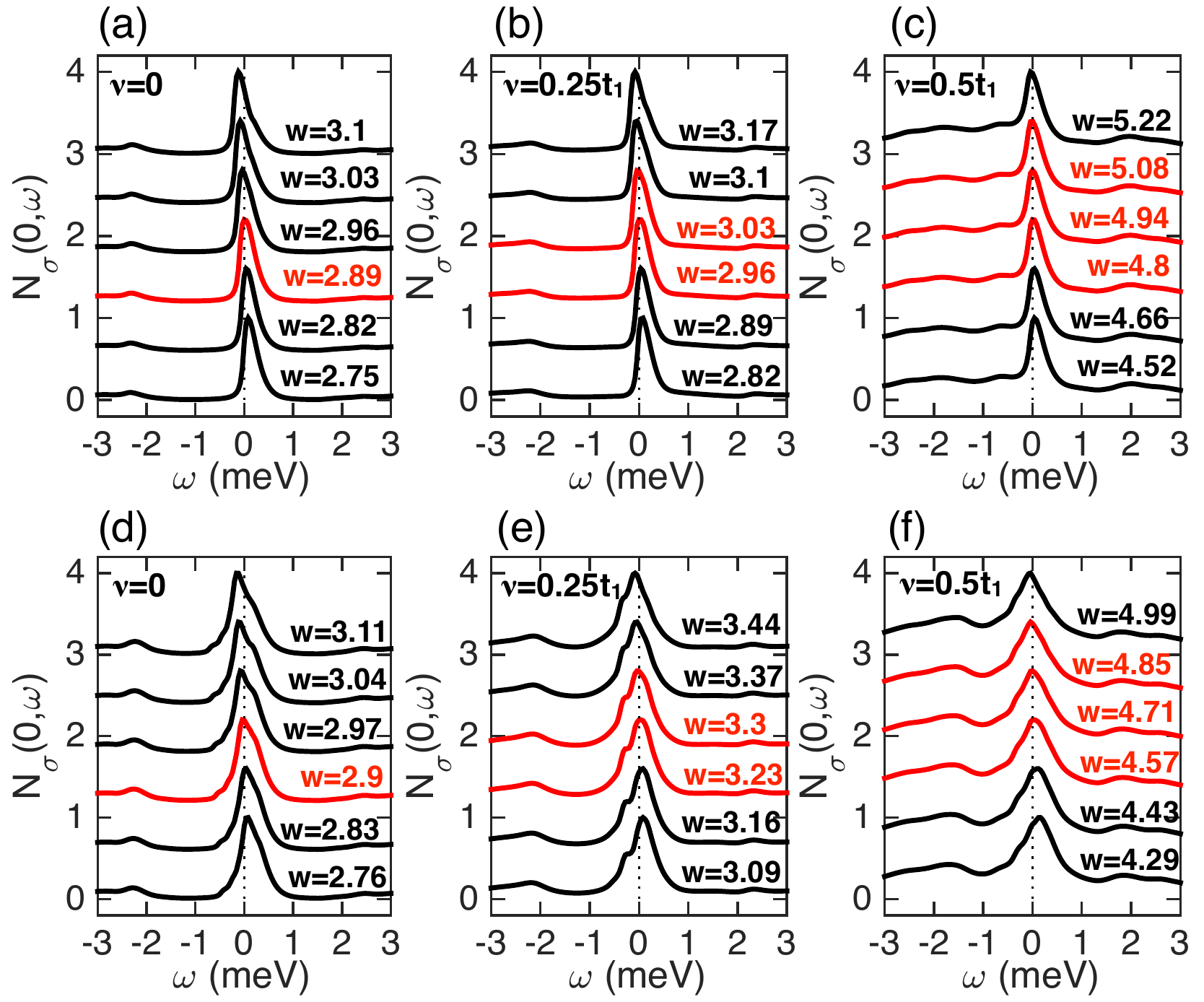}
\caption{
SOC effects on the ZBP for $s_{+-}$ pairing at $h_\text{ext}=0$ for a single impurity (top) and for 8 impurities in a square shape (bottom) as in Fig.~\ref{fig:fig3}. For $\nu=0$, (a) and (d), the ZBP appears at the critical value of impurity strength $w=w_c(\nu=0)$. For finite values of $\nu$, (b), (c), (e), and (f), the ZBP remains in a range of impurity strengths ($\Delta w=w_U-w_L$) and the robustness increases with increasing SOC. Note that $\Delta w$ for a single impurity is the same as those for multiple impurities at the finite values of the SOC strength $\nu$. 
}
\label{fig:fig4}
\end{figure}
%============================================

For numerical analysis, we consider 8 impurities arranged in the shape of a square  with the nearest neighbors separated by $d/2 = 6$ lattice sites (various other arrangements of impurities give qualitatively the same results). For this multiple-impurity problem, the Green's function, Eq.~(\ref{green_func}), is modified to include the scattering from the neighboring impurities $\mathbf{r}_I$:
\begin{eqnarray}
	\hat G(\mathbf{r},\mathbf{r};\omega)&=&\hat G^{(0)}(\mathbf{0},\omega)\\\nonumber
	&&+\sum_{I,I'}\hat G^{(0)}(\delta\mathbf{r}_I,\omega) \hat T(\mathbf{r}_I,\mathbf{r}_{I'},\omega) \hat G^{(0)}(-\delta\mathbf{r}_{I'},\omega),
\end{eqnarray}
where $\delta\mathbf{r}_I=\mathbf{r}-\mathbf{r}_I$ runs for all impurities in the system, and $\hat{T}(\mathbf{r}_I,\mathbf{r}_{I'},\omega)=[\hat I - \hat V G^{(0)}(\mathbf{r}_I-\mathbf{r}_{I'},\omega)]^{-1} \hat V$ is the 8$\times n_\text{imp}$-matrix with $n_\text{imp}$ being the number of the impurities in the system. In Fig.~\ref{fig:fig3}, we present the spin-resolved LDOS $N_\sigma(\mathbf{0},\omega)$ at one of the impurity sites ($\mathbf{r}_I=\mathbf{0}$) for clean system (green and magenta lines) and the impurity induced bound states at the critical scattering $w_c$ (blue and red lines) with and without SOC. The Zeeman splitting in the absence of SOC is manifest in the shift of each spin component in the LDOS for the clean system with increasing Zeeman field, as shown in Fig.~\ref{fig:fig3}(b) and (c). As the Zeeman field increases, the degeneracy between the two spin components is removed, thus $N_\uparrow$ and $N_\downarrow$ split from each other. On the other hand, even weak SOC ($\nu= 0.25 t_1$) dramatically changes the zero energy bound states, maintaining the ZBS even in the presence of the applied magnetic field $\sim 8$ Tesla. Fig.~\ref{fig:fig3}(d) shows the appearance of the ZPBs at the critical value of the impurity scattering $w=w_c(\nu)$ in the presence of SOC $\nu = 0.25 t_1$ with no magnetic field applied.  Fig.~\ref{fig:fig3} (e) and (f) illustrate the robustness of the ZBSs to the magnetic field $h_\text{ext}$.

In addition to robustness to applied magnetic field, in the presence of SOC and low energy quasi-particle states within the gap, the ZBSs in $s_{+-}$-wave superconductors also become robust to variations in the magnitude of the impurity scattering potential. This is important because without SOC the appearance of the ZBS at magnetic impurity sites in $s_{+-}$ superconductors requires fine tuned scattering potential at criticality $w_c(\nu=0)$ \cite{Tsai-2009}. In contrast, SOC enforces the appearance of ZBS in a range of magnetic impurity scattering potential above the critical value $w=w_c(\nu)$. Fig.~\ref{fig:fig4}(b), (c), (e), and (f) show that the presence of SOC facilitates the emergence of ZBSs in a broad range of the impurity strength, whereas in the absence of SOC, a ZBS is allowed only at the critical impurity scattering $w=w_c(\nu)$ [Fig.~\ref{fig:fig4}(a) and (d)]. In this plot, to illustrate the impurity strength dependence of the peak positions, $N_\sigma $ is normalized by the height as $\tilde N_\sigma = N_\sigma / \text{max}(|N_\sigma|)$.

%============================================
\begin{figure}[t]
\includegraphics[width=1.0\linewidth]{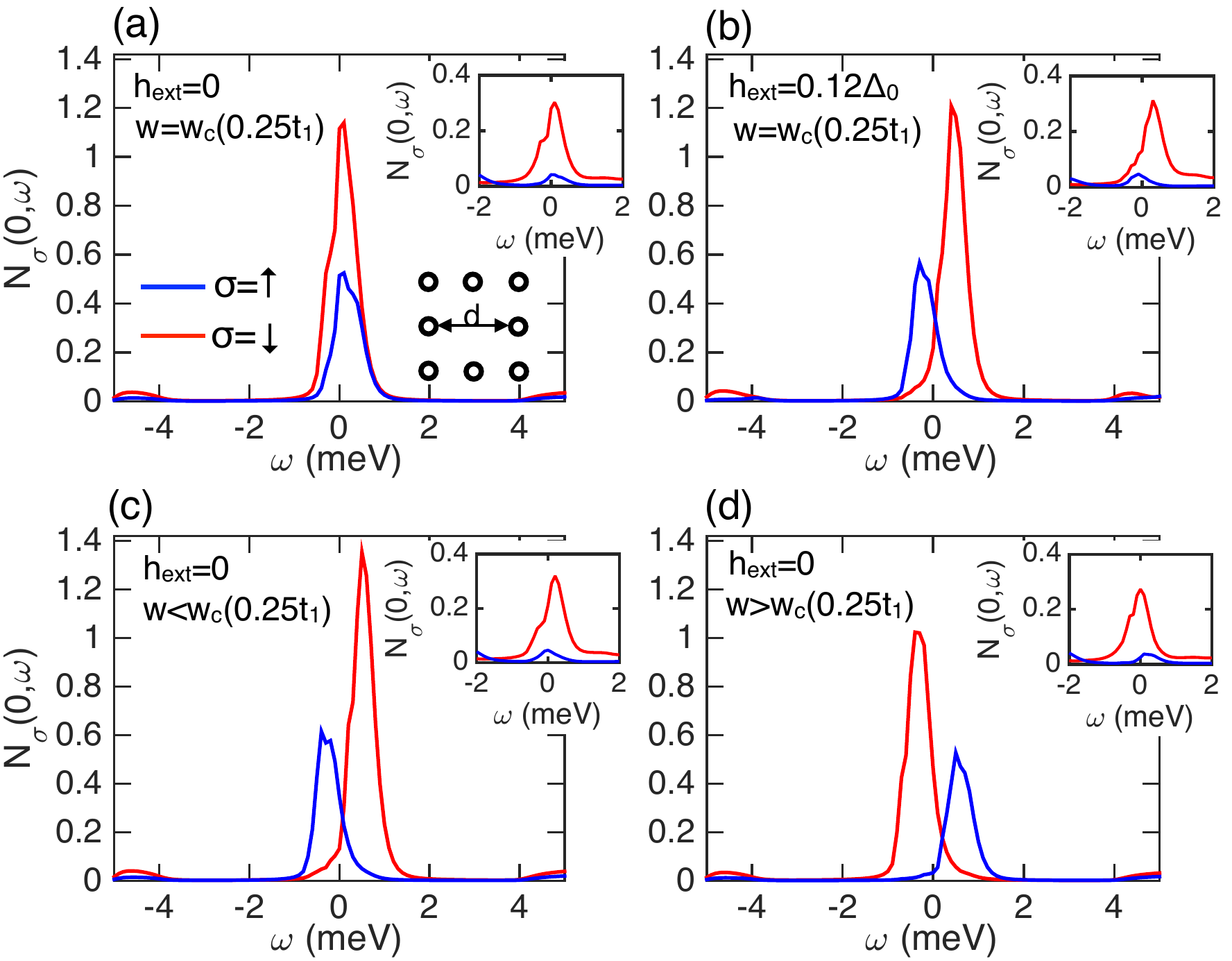}
\caption{
Effects of magnetic field and impurity strength on the in-gap states for $s_{++}$ pairing induced by 8 magnetic impurities in the presence of SOC. (a) At $w_c$, the peaks for $\sigma=\uparrow$ and $\downarrow$ are located at zero bias $\omega=0$ for $h_\text{ext}=0$. (b) Magnetic field splits into double peaks even in the presence of SOC ($\nu=0.25t_1$). (c) and (d) illustrate the ZBPs are sensitive to values of impurity strengths at $\nu=0.25t_1$. Insets illustrate the robustness of the ZBPs for $s_{+-}$ pairing against the magnetic field and impurity strengths at the same SOC strength, $\nu=0.25 t_1$.
}
\label{fig:fig5}
\end{figure}
%============================================

In contrast to unconventional sign-changing $s_{+-}$ pairing superconductors, magnetic impurity-induced ZBSs in conventional $s_{++}$ pairing ($s$-wave gap of the same sign on both hole and electron pockets) is \textit{not} robust to applied magnetic fields even in the presence of SOC. Fig.~\ref{fig:fig5} shows the spin-resolved LDOS $N_\sigma$ at a given impurity site ($\mathbf{r}_I=0$) for a spin-orbit coupled superconductor ($\nu=0.25t_1$) with $s_{++}$ pairing order parameter ($\Delta_\mathbf{k} = \Delta_0$) for various values of applied magnetic field and the magnitudes of impurity strengths. Fig.~\ref{fig:fig5} (a) shows the appearance of the ZBS at $w=w_c(\nu=0.25t_1)$ at zero applied magnetic field and (b) shows the pronounced Zeeman splitting of the peaks even in the presence of SOC, which is in stark contrast to the case of $s_{+-}$ pairing as shown in the insets. It is noteworthy that the ZBSs for the $s_{++}$ pairing requires fine tuning in the impurity potential even in the presence of SOC. Fig.~\ref{fig:fig5} (c) and (d) show the double peak structure for $\sigma=\uparrow$ and $\sigma=\downarrow$ for $w<w_c(\nu=0.25)$ and $w>w_c(\nu=0.25)$, respectively, while the insets show the robustness of the ZBSs with SOC to variations in the magnitude of the impurity potentials for $s_{+-}$ symmetry of the order parameter.

\section{Discussion and Conclusion}\label{sec4}
 The formation of localized sub-gap bound states at non-magnetic and magnetic impurity sites in unconventional ($s_{+-}$) and conventional ($s_{++}$) multi-band superconductors have been studied earlier \cite{Tsai-2009,Bang-2009,Li-2009,Akbari-2010,Ng-2009,Matsumoto-2009,Zhang-2009,Kariyado-2010,Beaird-2012}. Similar to the case of Yu-Shiba-Rusinov states \cite{Yu-1965,Shiba-1968,Rusinov-1969} in spin-singlet single-band superconductors \cite{Balatsky-2006,Yazdani-1997,Balatsky-1995}, it was found earlier that magnetic impurities induce localized zero energy bound states in multi-band $s_{\pm}$ and $s_{++}$ superconductors. Such zero energy bound states, however, occur at a critical value of the impurity scattering potential and are unlikely to be observed in experiments without fine tuning. Moreover, in analogy to their counterparts in single-band superconductors, the YSR states localized at magnetic impurities in multi-band superconductors are also unstable to applied Zeeman fields. Applied magnetic fields, therefore, are expected to split the magnetic impurity induced STM zero bias conductance peaks, if any, in a double peak structure. In recent experiments \cite{Yin2015}, however, robust STM peaks have been observed in a class of iron based superconductors, which remain unsplit even by a magnetic field as high as $\sim 8 T$. Concurrently, in another set of experiments \cite{Borisenko2016}, a substantial spin-orbit coupling $\sim 5-10 meV$ have been observed in all the classes of iron based superconductors in high resolution ARPES experiments. This has prompted us to investigate the effects of spin-orbit coupling on the Yu-Shiba-Rusinov states induced at magnetic impurities in multi-band $s_{\pm}$ and $s_{++}$ superconductors. Using a numerical T-matrix formalism and supporting theoretical arguments we have shown that robust zero energy bound states (that remain unsplit even by a magnetic field as high as $\sim 8 T$) are induced at isolated magnetic impurity sites in multi-band unconventional sign-changed $s_{+-}$ superconductors in the presence of spin-orbit coupling. No such enhancement of robustness by the effects of spin-orbit coupling is present for magnetic impurities in conventional $s_{++}$ multi-band superconductors.   
 
 The robustness of magnetic impurity induced zero bias states in sign-changing $s_{\pm}$ superconductors to variations in the impurity scattering potentials as well as the applied magnetic fields are the consequences of spin-orbit coupling along with low energy quasiparticle states within the superconducting gap. As we have shown, spin-orbit coupling (and also the presence of nearby magnetic impurities) effectively reduces the magnitude of the superconducting gap in $s_{\pm}$ (but not in $s_{++}$) superconductors. Since the impurity induced YSR states are pinned to the sub-gap energies even in the presence of a Zeeman field (see Appendix), the reduction of the superconducting gap (on one of the Fermi surfaces) in $s_{\pm}$ superconductors by spin-orbit coupling effectively ensures that the YSR states will remain pinned to zero or low energies in $s_{\pm}$ superconductors even in the presence of a substantial magnetic field. As no reduction of the superconducting gap occurs in $s_{++}$ superconductors by the effects of spin-orbit coupling, STM zero bias conductance peaks from magnetic impurities in $s_{++}$ superconductors remain strongly split by magnetic fields even in the presence of spin-orbit coupling. These simple and intuitive arguments, supported by theoretical and numerical evidence presented in this paper, provides one possible explanation of the recent observation \cite{Yin2015} of robust STM zero bias peaks at isolated magnetic impurities in one class of iron-based superconductors, without having to invoke exotic physics such as topological superconductivity \cite{Tai2015,Yin2015} and/or the absence of exchange coupling of magnetic impurities with the underlying Fe lattice \cite{Huang2016}. It is important to reiterate that we find the zero energy bound states localized at magnetic impurity sites in conventional sign-unchanged ($s_{++}$) superconductors to be strongly sensitive to applied magnetic fields and variations in the impurity potentials even in the presence of spin-orbit coupling. Thus, our results, in addition to filling the gap of analyzing the effects of spin-orbit coupling on YSR states in multi-band superconductors and providing one possible theoretical explanation of the observation of robust zero bias peaks in iron based superconductors, may help identify the order parameter symmetry of these superconductors as sign-changing $s_{+-}$ wave.

\begin{acknowledgments}
K.S and S.T are supported by AFOSR
(FA9550-13-1-0045). J.D.S would like to acknowledge
the University of Maryland, Condensed Matter theory
center, and the Joint Quantum institute for startup support.
\end{acknowledgments}

\appendix{}
\section{Subgap states for weak coupling}
Here we show that under quite generic circumstances there is always a sub-gap state bound to magnetic impurities in two dimensional BCS superconductor with an isotropic gap. The bound states to an impurity in a lattice are obtained by solving for the poles of the T-matrix or equivalently the zeros of an effective Hamiltonian that is written as
\begin{align}
&H_\text{eff}(\omega)=V-G^{(0)-1}(\omega)\label{Heff}
\end{align}
where in the case of a single-lattice site impurity, $G^{(0)}(\omega)$ is calculated on one lattice site. The single-site Green function ($G^{(0)}(\omega)$ ) can be expanded in terms of BdG eigenstates of the bulk as
\begin{align}
&G^{(0)}(\omega)=\sum_{n,\bf{k}}\frac{\Psi_{n\bf{k}}\Psi^\dagger_{n\bf{k}}}{\omega-\epsilon_{n\bf{k}}}=\int d\epsilon \frac{\rho_0(\epsilon)}{\omega-\epsilon}\chi(\epsilon),\label{G0}
\end{align}
where $\rho_0(\epsilon)$ is the DOS of BdG quasiparticles and $\chi(\epsilon)=\rho_0(\epsilon)^{-1}\sum_{n,\bf{k}}\Psi_{n\bf{k}}\Psi^\dagger_{n\bf{k}}\delta(\epsilon-\epsilon_{n\bf{k}})$.

Following Ref.~{\onlinecite{Brydon}}, we note that in the presence of rotational symmetry about the $z$-direction the Green function  conserves spin (i.e. $[G^{(0)}(\omega),\sigma_z]=0$), so that it commutes with the impurity Hamiltonian $V$.  In this case the two operators in the effective Hamiltonian in Eq.~(\ref{Heff}) can be simultaneously diagonalized and a bound state occurs whenever $V=\lambda^{-1}$, for some eigenvalue $\lambda$ of $G^{(0)}(\omega)$.  In this appendix we focus on weak impurity strengths, $V$, which could produce states near the gap edge $\Delta_1$. For 2D BCS superconductors with a rotationally symmetric gap,  $\rho_0(\epsilon)$ diverges near the gap edge  as $\epsilon\rightarrow \Delta_1$ and thus one can approximate the energy dependence of $\chi(\epsilon)$ in Eq.~(\ref{G0}) as $\chi(\epsilon)\approx \chi_0+\epsilon\chi_1$. Therefore, we can ignore the energy dependence of $\chi(\epsilon)\approx \chi_0$ in Eq.~(\ref{G0}) and conclude that in the limit $\omega\rightarrow \Delta_1$, the eigenvalue of $G^{(0)}(\omega)$ can be approximated as $\lambda\approx f(\omega) \chi_{0,n}$, where $f(\omega)=\int d\epsilon \frac{\rho_0(\epsilon)}{\omega-\epsilon}$  and $\chi_{0,n}$ are eigenvalues of $\chi_0$.

Since  the integral $f(\omega)$ is divergent as $\omega\rightarrow \Delta_1$ and $\chi_0$ has at least one finite (non-zero) eigenvalue, $\lambda^{-1}\propto f(\omega)^{-1}\rightarrow 0$ as $\omega\rightarrow \Delta_1$. Therefore, the impurity strength, $V=\lambda^{-1}$, required to produce a bound state near the gap edge (i.e. at $\omega\sim \Delta_1$) vanishes as the bound state energy $\omega$ approaches the gap edge. This implies that there is a bound state at arbitrarily small impurity strengths inside the spectral gap independent of how far the gap is suppressed.

%\bibliography{}

%\appendix{}

\end{document}